\documentclass[10pt]{iopart}
\usepackage{epsfig}
\usepackage{isolatin1}

\newcommand{\myscalebox}[1]{\scalebox{0.5}{#1}}
\newcommand{\myscaleboxb}[1]{\scalebox{0.4}{#1}}
\newenvironment{myalgorithm}[1]
{
\begin{tabbing} xx \= xx \= xx \= xx \= xx \= xx \= xx \= xx \kill
 {\bf algorithm #1}\\
 {\bf begin}\\
}
{
 {\bf end}\\
 \end{tabbing}
}

\begin{document}
\title{Statistical Mechanics of the vertex-cover problem}

\author{Alexander K. Hartmann and Martin Weigt}
\address{Institute for Theoretical Physics,
University of G\"ottingen, Bunsenstr. 9, 37073 G\"ottingen, Germany}
\eads{
\mailto{hartmann@theorie.physik.uni-goettingen.de},
\mailto{weigt@theorie.physik.uni-goettingen.de}}

\begin{abstract}
  We review recent progress in the study of the vertex-cover problem
  (VC). VC belongs to the class of NP-complete graph theoretical
  problems, which plays a central role in theoretical computer
  science. On ensembles of random graphs, VC exhibits an
  coverable-uncoverable phase transition. Very close to this
  transition, depending on the solution algorithm, easy-hard
  transitions in the typical running time of the algorithms occur.
  
  We explain a statistical mechanics approach, which works by mapping
  VC to a hard-core lattice gas, and then applying techniques like the
  replica trick or the cavity approach. Using these methods, the phase
  diagram of VC could be obtained exactly for connectivities $c<e$,
  where VC is replica symmetric. Recently, this result could be
  confirmed using traditional mathematical techniques.  For $c>e$, the
  solution of VC exhibits full replica symmetry breaking.
  
  The statistical mechanics approach can also be used to study
  analytically the typical running time of simple complete and
  incomplete algorithms for VC. Finally, we describe recent results
  for VC when studied on other ensembles of finite- and
  infinite-dimensional graphs.
\end{abstract}

\submitto{\JPA}

\maketitle

\section{Introduction}

Starting in the 80s of the last century, there are growing relations
between the fields of statistical physics and (theoretical) computer
science. This is true in particular for the study of disordered glassy
systems in physics and the research on optimization problems in
computer science \cite{hemmen1985}. Both fields can profit strongly
from each other.  In one way computer science helps physics: Recently
developed efficient optimization algorithms \cite{opt-phys2001} help
to study the low-temperature behavior of physical models, like spin
glasses, random field systems or solid-on-solid models. On the other
hand also developments in statistical physics have helped to develop
or improve existing optimization algorithms.  The most prominent
example is the invention of the simulated annealing method
\cite{kirkpatrick1983}, which has been applied to a variety of
optimization problems.

In recent years another variant of how physics can help computer
science has emerged. Computational problems can be sorted into
different classes. From the viewpoint of a person wanting to solve
problems, a very convenient class is the class P: It collects all
problems which can be solved on a computer in a running time, which
grows even in the worst case only polynomially with the size of the
problem.  These problems are called {\em easy}.  In theoretical
computer science \cite{lewis1981,papadimitriou1994,sipser1997} these
problems are analyzed using model computers, e.g. the {\em Turing
  machine} (TM) \cite{turing1936}.  A {\em deterministic} TM can solve
the same problems like a conventional (von Neumann) computer.  But not
all problems can be solved polynomially. There are problems, for which
for sure no polynomial algorithm exists. These problems are called
{\em hard}.  But most of these problems have only academic
applications.  The most interesting problems lie on the interface
between polynomial and exponential running time. They belong to the
class of {\em nondeterministic polynomial} problems (NP)
\cite{garey1979}.  This means that a {\em nondeterministic} TM can
solve any problem from NP in polynomial time. This works in the
following way: First, the nondeterministic abilities of the TM are
used to generate a solution. Then the TM proves deterministically that
the solution is correct.  For purely deterministic computers, all
algorithms for solving problems from NP known so far need in the worst
case an exponentially growing running time. Hence, it appears that the
problems from NP are hard as well.  But so far there is no {\em proof}
that the problems from NP are indeed hard. This is the so called {\em
  P-NP problem}, one of the great open questions in computer science
\footnote{The Clay Mathematics Institute of Cambridge, Massachusetts
  (CMI) has designated a \${}1 million price for the solution of the
  P-NP problem.}. Expressed in colloquial language we have to answer
the question: ``What is it that makes a problem hard ?''

A notable advance \cite{review1,review2} towards the answer of this
question has recently been achieved by realizing that worst case and
typical case are different. This means that for some problems there
are ensembles of problems which can be solved typically in polynomial
time, while the worst case is still exponential. In particular, there
are suitably parametrized ensembles of random problem instances, where
in one region of parameter space the instances are easy while in
another region the instances are hard \cite{mitchell1992,selman1994}.
The typically hardest to solve instances are often found {\em at} the
boundaries separating these regions.  The effects found at the
boundaries have much in common with phase transitions in physical
systems \cite{yeomans1992,goldenfeld1992}. Recently methods from
statistical physics \cite{mezard1987}, like the replica trick or the
cavity approach, have been applied to classical problems from computer
science. The most prominent one is the satisfiability problem (SAT)
\cite{garey1979}.  SAT is the most famous and central of all problems
in theoretical computer science: In 1971, it was the first one which
was shown  by Cook \cite{cook1971} to be {\em NP-complete}, which means
that all problems from NP can be mapped onto SAT using polynomial
algorithms. Hence, SAT is at least as hard as any problem in NP. Using
the statistical mechanics approach it is possible to obtain results
which have not been found before using classical mathematical methods
\cite{monasson1997,monasson1999,brioli2000,ricci2001}.  Furthermore 
this approach allows to invent new algorithms which are sometimes substantially
faster than previously know algorithms \cite{mezard2002}.

In this paper, we review the recent progress in the field by
concentrating on the vertex-cover problem (VC), which belongs to the
six ``classical'' NP-complete problems in theoretical computer science
\cite{garey1979}.  VC is a problem defined on graphs. We first
introduce VC and show that it is NP-complete. Then we present some
algorithms which can be used to solve NP. In the succeeding section,
we present results characterizing
 the phase transition, which occurs when studying
VC on ensembles of random graphs. Next, we describe the results
obtained for the phase diagram using statistical mechanics methods. In
section six we show how the typical running time of algorithms can be
analyzed analytically. Next, we consider other ensembles of random
graphs, especially scale-free graphs and graphs consisting of a
collection of connected cliques.  Finally, we summarize and give an
outlook.

\section{The vertex-cover problem}

In this section, we will introduce the terminology, show that VC is
NP-complete and review some rigorous results about vertex cover which
have been obtained previously by applying mathematical techniques.

\subsection{Vertex cover and related problems}\label{sec:vc}

Let us start with the definition of vertex covers. We consider a graph
$G=(V,E)$ with $N$ vertices $i\in \{1,2,...,N\}$ and undirected edges 
$\{i,j\}\in E\subset V\times V$ connecting pairs of vertices. Please
note that $\{i,j\}$ and $\{j,i\}$ both denote the same edge.

Definition 1: {\it A vertex cover $V_{vc}$ is a subset $V_{vc}\subset
  V$ of vertices such that for all edges $\{i,j\}\in E$ at least one of
the endpoints is in $V_{vc}$, i.e. $i\in V_{vc}$ or $j\in V_{vc}$.}

Later on also subsets $V^\prime$ are considered, which are not
covers. Anyway, we call all vertices in $V^\prime$ {\em covered}, all
others {\em uncovered}. Also edges from $E\cap ([V^\prime\times V] \cup[ V
\times V^\prime])$ are called covered.  This means that  $V^\prime$ is a
vertex cover, iff all edges are covered.

There are three different variants of VC:

\begin{itemize}
\item[P1] The {\em minimal vertex-cover} problem, which consists in
  finding a vertex cover $V_{vc}$ of minimal cardinality, and
  calculate the minimal fraction $x_{\rm c}(G)=|V_{vc}|/N$ needed to
  cover the whole graph.
\item[P2] The {\em decision variant} of this problem is: ``Given a
  number $X=xN$, is there a vertex cover $V_{vc}$ of size $X$?''.
\item[P3] If there is no vertex cover of size $X$, one can study the
  related {\em optimization problem}: Find a set $V^\prime$ with
  $|V^\prime|=X$ which minimizes the number of uncovered edges. In
  other words, we try to distribute $X$ covering marks on the $N$
  vertices in an optimal way, such that the following {\it energy} of 
  configurations is minimized:
  \begin{equation}  
    \hspace*{-1.8cm}
    E(G,x)=\min \{ \mbox{number of {\em uncovered}\/ edges when
      covering } xN \ 
    \mbox{vertices} \} \label{eq:coverEnergy}
  \end{equation}
  This means, the graph is coverable using $X=xN$ vertices iff the
  ground state energy is zero.
\end{itemize}

VC is equivalent to other  problems:
\begin{itemize}
\item An {\it independent set} is a subset of vertices which are
  pairwise disconnected in the graph $G$. Due to the above-mentioned
  properties, any set $V\setminus V_{vc}$ thus forms an independent
  set, and maximal independent sets are complementary to minimal
  vertex covers.
\item A {\it clique} is a fully connected subset of vertices, and thus
  an independent set in the complementary graph $\overline{G}$ where
  vertices $i$ and $j$ are connected whenever $\{i,j\}\notin E$ and
  vice versa.
\end{itemize}

\subsection{NP-completeness}

Here, we show the NP-completeness of VC \cite{garey1979}.  For this
purpose, we first introduce the 3-satisfiability problem (3-SAT),
which is know to be NP-complete. Then we show how 3-SAT can be mapped
onto VC in polynomial time.

3-SAT is a problem concerning Boolean formulas.  A Boolean formula $F$
in $K=3$ {\em conjunctive normal form} (CNF) has the following
structure: It is a formula over $N$ boolean variables
$\{x_1,x_2,\ldots,x_N\}$ which contains $M$ {\em clauses} $C_i$:
$F=C_1\wedge C_2\wedge \ldots \wedge C_M$.  Each clause is a
disjunction of three literals $C_p=l_p^1\vee l_p^2\vee l_p^3$, where
each literal is either a variable ($l^i_p=x_j$) or a negated variable
($l^i_p=\overline{x_j}$). The 3-SAT problem is: 
\begin{center}
``Given a 3-CNF formula
$F$, is there an assignment of the variables
$\{x_1,\ldots,x_N\}\in\{0,1\}^N$ such that $F$ evaluates to {\em
  true}, i.e., is $F$ {\em satisfiable}? `` 
\end{center}
3-SAT is a special variant of
SAT and has been proven to be NP-complete before \cite{garey1979}.
The proof of the NP-completeness of VC works by reducing 3-SAT to VC
in polynomial time.

First, we show VC$\in$NP: It is very easy to decide for a given subset
$V^\prime$ of vertices, whether all edges are covered, i.e. whether
$V^\prime$ is a vertex cover, by just iterating over all edges.

Hence, it remains to show that 3-SAT is polynomially reducible to VC,
(one writes 3-SAT $\le_p$ VC).

Let $F=C_1\wedge \ldots\wedge C_m$ be a 3-SAT formula with variables
$X=\{x_1,\ldots, x_n\}$ and $C_p=l_p^1\vee l_p^2 \vee l_p^3$ for all
$p$.

We have to create a graph $G$ and a threshold $K$, such that $G$ has a
VC of size lower than or equal to $K$, iff $F$ is satisfiable. For
this purpose, we set:
\begin{itemize}
\item $V_1\equiv \{v_1,\overline{v}_1, \ldots, v_n, \overline{v}_n\}$,
  ($|V_1|=2n$) and $E_1=\{
  \{v_1,\overline{v}_1\},\{v_2,\overline{v}_2\}, \ldots,
  \{v_n,\overline{v}_n\}\}$, i.e. for each variable occurring in $F$
  we create a pair of vertices and an edge connecting it.
  
  To cover the edges in $E_1$, we have to include at least one vertex
  per pair in the covering set. In this part of the graph, each cover
  corresponds to an assignment of the variables with the following idea
  behind it: If for variable $x_i=1$, then $v_i$ should be covered,
  while if $x_i=0$ then $\overline{v}_i$ is to be covered. It will
  become clear soon, why this correspondence has been chosen.
\item For each clause in $F$ we introduce three vertices connected in
  form of a triangle: $V_2\equiv \{
  a_1^1,a_1^2,a_1^3,a_2^1,a_2^2,a_2^3,\ldots a_m^1,a_m^2,a_m^3\}$ and
  $E_2 =\{ \{a_1^1,a_1^2\}$, $ \{a_1^2,a_1^3\}$, $ \{a_1^3,a_1^1\}$, $
  \{a_2^1,a_2^2\}$, $ \{a_2^2,a_2^3\}$, $ \{a_2^3,a_2^1\}$, $ \ldots,
  \{a_n^1,a_n^2\}$, $ \{a_n^2,a_n^3\}$, $ \{a_n^3,a_n^1\} \}$,
  
  Per triangle, i.e. per clause, we have to include at least two
  vertices in a VC. We intent that in a 
  cover of minimum size, the {\em uncovered}
  vertex corresponds to a literal which is satisfied. This will be
  induced by the edges generated in the following.
  
\item Finally, for each position $i$ in a clause $p$, vertex $a_p^i$
  is connected with the vertex representing the literal $l_p^i$
  appearing at that position of the clause: $E_3\equiv $ $\{
  \{a_p^i,v_j\} | p=1,\ldots,m,\, i=1,2,3\, \mbox{ if } l_p^i=x_j\}
  \cup \{ \{a_p^i,\overline{v}_j\} | p=1,\ldots,m,\, i=1,2,3\, \mbox{
    if } l_p^i=\overline{x}_j\}$. Hence, $E_3$ contains edges each
  connecting one vertex from $V_1$ with one vertex from $V_2$.
\item The graph $G$ is the combination of the above introduced
  vertices and edges: $G=(V,E)$, $V=V_1\cup V_2$, $E=E_1\cup E_2 \cup
  E_3$.
\item The size of the vertex cover to be constructed is set to
  $K\equiv n+2m$.
\end{itemize}

\begin{figure}[ht]
\begin{center}
\scalebox{0.4}{\includegraphics{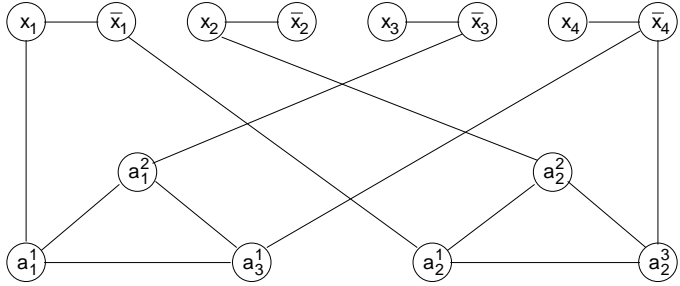}}
\caption{VC instance resulting from the 3-SAT instance
$F=(x_1 \vee \overline{x}_3 \vee
\overline{x}_4)$  $\wedge$  $(\overline{x}_1 \vee x_2 \vee
\overline{x}_4)$. }
\label{figVCgraph}
\end{center}
\end{figure}

In the following example, we show how the transformation works for a
small 3-SAT formula:

{\bf Example} We consider $F=(x_1 \vee \overline{x}_3 \vee
\overline{x}_4)$ $\wedge$ $(\overline{x}_1 \vee x_2, \overline{x}_4)$.
The resulting graph $G(V,E)$ is displayed in Fig. \ref{figVCgraph}

The number of vertices generated by this transformation   is $O(n+m)$,
i.e. linear in the sum of the number of clauses and the number of
variables of $F$. Since the number of variables is bounded by three
times the number clauses, the construction of the graph is linear in
the length of $F$, i.e. in particular polynomial.  It remains to show:
$F$ satisfiable if and only if there exists a vertex cover $V^\prime$
of $G$ with size $|V^\prime|\le K$.

Now let $F$ be satisfiable and $\{X_i\}, X_i=0,1$ a satisfying
assignment.  We set $V_1^{\prime} = \{v_i | X_i=1\}\cup
\{\overline{v}_i| X_i=0\}$.  Obviously $|V_1^{\prime}|=n$ and all
edges in $E_1$ are covered.  For each clause $C_p$, since it is
satisfied by $\{X_i\}$, there is one satisfied literal $l_p^{i(p)}$.
We set $V_2^{\prime}=\{a_p^i | p=1,\ldots,m;\, i\neq i(p) \}$. We have
included 2 vertices per clause in $V_2$ (by excluding $a_p^{i(p)}$),
i.e. 2 vertices per triangle in $E_2$.  Thus, $|V_2^{\prime}|=2m$ and
all edges of $E_2$ are covered. Furthermore, since $l_p^{i(p)}$ is
satisfied, the vertex corresponding to the literal is in $V_1$, hence
all edges contained in $E_3$ are covered as well. To summarize
$V^{\prime} =V_1^{\prime} \cup V_2^{\prime}$ is a VC of $G$ and
$|V^{\prime}|=n+2m\le K$.

Conversely, let be $V^{\prime}\subset V$ be a VC of $G$ and
$|V^{\prime}|\le K$. Since a VC must include at least one vertex per
edge from $E_1$ and at least two vertices per triangle from $E_2$, we
know $|V^{\prime}|\ge n+2m=K$, hence we have $|V^{\prime}|=K$, i.e.
{\em exactly one} vertex per pair $x_i,\overline{x_i}$ and {\em
  exactly two} vertices per triplet $a_p^1,a_p^2,a_p^3$ is included in
$V^{\prime}$. Now we set $X_i=1$ if $x_i\in V^{\prime}$ and $X_i=0$ if
$x_i\not\in V^{\prime}$. Since each triangle (each corresponding to a
clause), has one vertex $a_p^i(p)\not\in V^{\prime}$, we know that the
vertex from $V_1$ connected with it is covered. Hence, the literal
corresponding to this vertex is satisfied. Therefore, for each clause,
we have a satisfied literal, hence $F$ is satisfied and $\{X_i\}$ is a
satisfying assignment.

\subsection{Vertex covers of random graphs}
\label{sec:bound}

In order to speak of median or average cases, and of phase
transitions, we have to introduce a probability distribution over
graphs. This can be done best by using the concept of {\it random
graphs} as already introduced about 40 years ago by Erd\"os and
R\'enyi \cite{ErRe}.  A random graph $G_{N,p}$ is a graph with $N$
vertices $V=\{1,...,N\}$, where any pair of vertices is connected randomly
and independently by an edge with probability $p$. So the expected
number of edges becomes $p {N\choose 2} = pN^2/2+O(N)$, and the
average connectivity of a vertex equals $p(N-1)$.

We are interested  in the large-$N$ limit of {\it finite-connectivity
  graphs}, where $p=c/N$ with constant $c$. Then the
average connectivity $c+O(N^{-1})$ stays finite. In this case, we also
expect the size of minimal vertex covers to depend only on $c$,
$x_{\rm c}(G)=x_{\rm c}(c)$ for almost all random graphs $G_{N,c/N}$.

Next we are going to present some previously derived rigorous bounds
on $x_{\rm c}(c)$. A general one for arbitrary, {\it i.e.} non-random
graphs $G$ was given by Harant \cite{Ha} who generalized an old result
of Caro and Wei \cite{CaWe}. Translated into our notation, he showed
that
\begin{equation}
  \label{bound_harant}
 x_{\rm c}(G)\leq 1-\frac{1}{N}\frac{\left(\sum_{i\in V}\frac{1}{d_i+1} 
                                   \right)^2}{
                  \sum_{i\in V}\frac{1}{d_i+1} - \sum_{(i,j)\in E} 
                      \frac{(d_i-d_j)^2}{(d_i+1)(d_j+1)}}
\end{equation}
where $d_i$ is the connectivity (or degree) of vertex $i$.  This can
easily be converted into an upper bound on $x_{\rm c}(c)$ which holds
almost surely for $N\to\infty$.

The vertex cover problem and the above-mentioned related problems were
also studied in the case of random graphs, and even completely solved
in the case of infinite connectivity graphs, where any edge is drawn
with finite probability $p$, such that the expected number of edges is
$p {N\choose 2}=0(N^2)$. There the minimal VC has cardinality
$(N-2\ln_{1/(1-p)}N-O(\ln \ln N))$ \cite{BoEr}. Bounds in the
finite-connectivity region of random graphs with $N$ vertices and $cN$
edges were given by Gazmuri \cite{Ga}. He has shown that
\begin{equation}
  \label{bound_gazmuri}
  x_l(c) < x_{\rm c}(c) < 1- \frac{\ln c}{c}
\end{equation}
where the lower bound is given by the unique solution of
\begin{equation}
  \label{low}
  0=  x_l(c) \ln  x_l(c) + (1- x_l(c)) \ln (1- x_l(c))
      - \frac{c}{2} (1- x_l(c))^2\ .
\end{equation}
This bound coincides with the so-called annealed bound in statistical
physics. The correct asymptotics for large $c$ was given by Frieze
\cite{Fr}:
\begin{equation}
  \label{asympt}
  x_{\rm c}(c) = 1 - \frac{2}{c}(\ln c - \ln\ln c +1 - \ln2 )
   +o\left(\frac{1}{c}\right)
\end{equation}
with corrections of $o(1/c)$ decaying faster than $1/c$.

Few studies have investigated VC on other ensembles of graphs. They
are reviewed in Sec. \ref{sec:otherEnsembles}.

\section{Algorithms}

There are two types of algorithms: incomplete and complete ones.
Complete algorithms guarantee to find the optimum or true solution,
hence the solution space is searched in principle completely. For
incomplete algorithms, it is not ensured that the true solution or the
global optimum is found. But they are very often sufficient for
practical applications.

\subsection{Incomplete Algorithms}
\label{sec:heuristics}

First, we present a greedy heuristic for finding small vertex covers,
i.e. approximation for the solutions of problem P1.  The basic idea of
the heuristic is to cover as many edges as possible by using as few
vertices as necessary. Thus, it is favorable to cover vertices with a
high degree. This step can be iterated, while the degree of the
vertices is adjusted dynamically by removing edges and vertices which
are covered.  This leads to the following algorithm, which returns an
approximation of the minimum vertex cover $V_{vc}$, the size
$|V_{vc}|$ is an upper bound of the true minimum vertex-cover size:
\begin{myalgorithm}{min-cover($G$)}
\> initialize $V_{vc}=\emptyset$;\\
\> {\bf while} there are uncovered edges {\bf do}\\
\> {\bf begin}\\
\>\> take one vertex $i$ with the largest current degree $d_i$;\\
\>\> mark $i$ as covered: $V_{vc} = V_{vc} \cup \{i\}$;\\
\>\> remove all incident edges $\{i,j\}$ from $E$;\\
\>\> remove vertex $i$ from $V$;\\
\> {\bf end};\\
\> return($V_{vc}$);\\
\end{myalgorithm}

It is easy to invent examples where the heuristic fails to find the
true minimum VC, e.g. a star graph having one center vertex 
to which $k>2$ arms of length 2 are attached.

This most simple heuristics has been generalized by one of the authors
within the framework of a random vertex selection \cite{weigt2002},
which is characterized by a parameter $k$ called {\em depth}. Each
vertex $i$ is selected with a probability $w_{d(i)}$ which depends on
the (current) degree $d(i)$ of the vertex.  Then, within the
generalized heuristic, a subgraph $G^{(k)}(i)=(V^{(k)}(i),E^{(k)}(i))$
is taken, where $V^{(k)}(i)$ contains all vertices which have at most
chemical distance $k$ from $i$. Here the chemical distance of two vertices
$j$ and $i$ counts the number of edges of the shortest path from $i$
to $j$. $E^{(k)}(i)$ contains the edges connecting the vertices from
$V^{(k)}(i)$. Then $G^{(k)}(i)$ is covered starting by covering all
vertices with distance $k$ from $i$ and then iteratively selecting
vertices $j$ among the remaining with maximal distance from $i$,
uncovering $j$ and covering all neighbors of $j$.  The results of an
analysis of the dynamics of this algorithm are reviewed in Sec.\ 
\ref{sec:generalizedHeuristic}.

The special case $k=1$ and $w_d=1$ has been analyzed by Gazmuri
\cite{Ga} for deriving the bound (\ref{bound_gazmuri}).  The greedy
heuristic presented before corresponds to the case $k=0$ and
$w_d=\delta_{d,d_{\rm max}}$, where $d_{\rm max}$ is the current
maximum degree in the graph. This case, where $w_d$ is dynamically
adjusted, have not be analyzed so far.

An alternative are incomplete algorithms based on conventional Monte
Carlo (MC) simulations in the grand-canonical ensemble, characterized
by a chemical potential $\mu$.  Here we present a variant \cite{hclg2003},
where one restricts the dynamics to true covers and allows movements
of the covering marks as well as fluctuations of the size of the
cover. First one selects an initial configuration, for example by
using the above heuristics or by covering all vertices. For each MC
step, a vertex $i$ is selected randomly. With probability $p$ (e.g.
$p=0.5$) a MOVE (M) step is performed, and with probability $1-p$ an
EXCHANGE (EX) step:
\begin{itemize}
\item[M] If vertex $i$ is covered and has exactly one uncovered
  neighbor, the covering mark is moved to the neighbor. In all other
  cases, the configuration remains unchanged.
\item[EX] If the site is uncovered, a covering mark is inserted with
  probability $\exp(-\mu)$. If the site is covered, and all neighboring
  sites are covered, the covering mark is removed from $i$.
\end{itemize}
Note that in this way detailed balance is fulfilled. Ground states,
i.e. minimum-size vertex covers can be obtained by starting with a
small chemical potential, which is slowly increased. The chemical
potential thus plays the same role in the algorithm as the decreasing
temperature in simulated annealing \cite{kirkpatrick1983}. Like the
latter algorithm, MC simulations can reach a
globally optimal vertex cover only on exponential time scales. 
On the other hand the Monte Carlo
approach allows to study dynamic properties of the model, which can be
regarded as a hard-core lattice gas \cite{hclg2003}, see also below.

The efficiency of randomized incomplete algorithms can be increased 
by introducing {\em restarts} \cite{gomes2000}.
The basic idea is to let the randomized
 algorithm run for a fixed number $\Delta T$ of steps. If
no solution is found in this time, the algorithm is restarted from the
beginning but with a different seed of the random number
generator. The basic idea behind this concept is that during a run the
system may be trapped in a local minimum, hence the chance of finding
a solution is increased when starting again.

\subsection{Complete Algorithms}
\label{sec:algorithms}

Next, we present two complete algorithms: They guarantee to find the 
exact answer, even if the time required will, in general, grow
exponentially with the graph size.

First we turn to the problem, where we are interested only in
minimum-size vertex covers (problem P1).  Since each vertex can be
either covered or uncovered, the most direct approach is to enumerate
all possible $2^N$ configuration, store all those being VCs, and
finally select one of those having minimal VC cardinality. Obviously,
the time-complexity of this approach is $O(2^N)$. Early attempts
\cite{balas1973,nemhauser1975} have the same worst-case running time.
The approach of Tarjan and Trojanowski \cite{tarjan77} presented here
has an $O(2^{N/2})$ time complexity. It uses a divide-and-conquer
approach.  First, all connected components of the graph are obtained.
Then the minimum-size vertex covers for all
components are calculated separately by recursive calls.  The treatment of each
connected component is based on the following idea.  Let $i\in V$ a
vertex, $A(i)\subset V$ its neighbors in $G$ and for any subset
$S\subset V$ let $G(S)=(S, E(S))$ the subgraph induced by $S$, i.e.
$E(S)=E\cap (S\times S)$.  Then the minimum-size vertex cover is
either $\{i\}$ combined with the minimum-size vertex cover of
$G(V\setminus\{i\})$ or $A(i)$ combined with the minimum-size vertex
cover of $G(V\setminus\{i\}\setminus A(i))$.

    Furthermore, the algorithm uses the concept of {\em
    domination}. This means basically that one considers small
    subgraphs $S$. Among all possible VCs of the subgraph one
    disregards all those, which provably cannot lead to better VCs of
    the full graph -- mainly because they cover only few or none of
    the edges connecting vertices from $S$ to $V\setminus S$.  We
    explain the simplest example for domination.  In this case  leaves
    are dominated,  i.e. vertices $i$ having only one single neighbor
    $j$. Here, for a minimum-size vertex cover one must cover either
    $i$ or $j$.  Since $i$ has only one neighbor, but $j$ may have
    more, we can immediately cover $j$ and remove the vertices $i,j$
    and all incident edges. This is the basic idea of the leaf-removal
    algorithm of Bauer and Golinelli \cite{bauer2001}. Note that this
    corresponds to the case depth $k=1$, $w_d=\delta_{d,1}$ of the
    generalized heuristic discussed in the last section.

The full algorithm is still deterministic but more general than leaf
removal: For each connected component, the vertex $i_0$ having the
smallest degree is determined. Degree $d_{i_0}=0$ corresponds to an
isolated vertex, which is not covered.  Degree $d_{i_0}=1$ corresponds
to a leave which is treated as discussed above. Furthermore, the
algorithm treats explicitly the cases of degree $d_{i_0}=$2,3 and 4.
For higher lowest degrees $d_{i_0}>4$, basically the subproblems for
$i_0$ covered and $i_0$ uncovered must be treated completely. But
during the recursive calls generated in this way, the cases with
smaller minimum degree might appear again.  The full detailed five
page presentation of the algorithm with all cases and subcases can be
found in Ref. \cite{tarjan77}. Due to the application of domination
the algorithm runs faster but it is unable to find more than {\em one}
minimal VC, hence it cannot be used to enumerate all solutions.

A simpler to implement algorithm \cite{Shindo90} exhibits a worse time
complexity $O(2^{n/2.863})$, but the authors claim that within their
computer experiments it was faster than the method of Tarjan and
Trojanowski.

If one is not only interested in one single minimum VC but in
enumerating all, the divide-and-conquer method does not work and
branch-and-bound approaches \cite{lawler66,lueling1992} must be
applied.  Also for the case where the number of covering marks $X$ is
given and one looks for all configuration of minimum energy (problem
P3), a branch-and-bound method is feasible. We will present an
algorithm for this latter case. The algorithm enumerating all
minimum-size VCs (problem P1) works in the same spirit.

The branch-and-bound approach differs from the previous method by the
fact that the concept of domination cannot be used. The basic idea is
to build the full configuration tree.  While doing this, the algorithm
makes certain choices where to put covering marks. If no VC of the
desired size is found, some covering marks have to be removed and to
be placed elsewhere, i.e. the algorithm has to backtrack. This is done
in a systematic way allowing to investigate the full configuration
space.  This $O(2^N)$ running time is reduced by omitting subtrees of
the full tree by using a {\em bound}: Trees where for sure no
minimum-energy configuration is located can be omitted. The bound
applied in the following algorithm uses the {\em current} vertex
degree $d(i)$, which is the number of uncovered neighbors at a
specific stage of the calculation. By covering a vertex $i$ the total
number of uncovered edges is reduced by exactly $d(i)$. If several
vertices $j_1,j_2,\ldots,j_k$ are covered, the number of uncovered
edges is {\em at most} reduced by $d(j_1)+d(j_2)+\ldots +d(j_k)$.
Assume that at a certain stage within the backtracking tree, there are
$uncov$ edges uncovered and still $k$ vertices to cover. Then a lower
bound $M$ for the minimum number of uncovered edges in the subtree is
given by
\begin{equation}
M=\max\left[0,uncov-\max_{j_1,\ldots,j_k}d(j_1)+\ldots+d(j_k)\right]\ .
\end{equation}
The algorithm can avoid branching into a subtree if $M$ is strictly
larger than the number $opt$ of uncovered edges in the best solution
found so far.  For the order the vertices are selected to be
(un-)covered within the algorithm, the following heuristic is applied:
the order of the vertices is given by their current degree. Thus, the
first descent into the tree is equivalent to the greedy heuristic
presented before. Later, it will be become clear from the results that
this heuristic is indeed not a bad strategy.

The following representation summarizes the algorithm for enumerating
all configurations exhibiting a minimum number of uncovered edges.
Let $G=(V,E)$ be a graph, $k$ the number of vertices to cover and
$uncov$ the number of edges to cover. Initially $k=X$ and $uncov=|E|$.
The variable $opt$ is initialized with $opt=|E|$ and contains the
minimum number of uncovered edges found so far. The value of $opt$ is
passed via call by reference.  At the beginning all vertices $i\in V$
are marked as {\it free}. The marks are considered to be passed via
call by reference as well (not shown explicitly).  Additionally it is
assumed that somewhere a set of (optimum) solutions can be stored.

\begin{myalgorithm}{min-cover($G,k,uncov,opt$)}
\> {\bf if} k=0 {\bf then} $\{$leaf of tree reached?$\}$\\
\> {\bf begin}\\
\>\> {\bf if} $uncov<opt$ {\bf then }$\{$new minimum found?$\}$\\
\>\> {\bf begin}\\
\>\>\> $opt:=uncov$;\\
\>\>\> clear set of stored configurations;\\
\>\> {\bf end};\\
\>\> store configuration;\\
\> {\bf end};\\
\> {\bf if} bound condition is true (see text) {\bf then}\\
\>\> {\bf return};\\
\> let $i\in V$ a vertex marked as {\it free } of maximal current degree;\\
\> mark $i$ as {\it covered};\\
\> $k:=k-1$;\\
\> adjust degrees of all neighbors $j$ of $i$: $d(j):=d(j)-1$;\\
\> {\bf min-cover}($G,k,uncov-d(i),opt$) $\{$branch into 'left' subtree$\}$;\\
\> mark $i$ as {\it uncovered};\\
\> $k:=k+1$;\\
\> (re)adjust degrees of all neighbors $j$ of $i$: $d(j):=d(j)+1$;\\
\> {\bf min-cover}($G,k,uncov,opt$) $\{$branch into 'right' subtree$\}$;\\
\> mark $i$ as {\it free};\\
\end{myalgorithm}

In the actual implementation, the algorithm does not descend further
into the tree as well, when no uncovered edges are left. In this case
the vertex covers of the corresponding subtree consist of the vertices
covered so far and all possible selections of $k$ vertices among all
uncovered vertices.

Finally we note that using the concepts of restarts one can also turn
a complete backtracking algorithm into a (possibly) faster incomplete
one. An application to VC has been studied by Montanari and Zecchina
\cite{montanari2002}. The algorithm must be randomized, for applying
restarts. Hence the choice which vertex is treated next is performed
in some random way, similar to the generalized heuristic presented
above. By applying many restarts, rare events become important: On one
hand, the latter may have exponentially smaller search trees, i.e. in
this case the algorithm by chance does not need to backtrack as long
as usually. On the other hand, events of this type are exponentially
rare. Balancing the exponential gain due to the smaller search tree
against the exponential loss due to large number of restarts required
to find such an event, an optimal backtracking (i.e. running) 
time per restart can be
found. The analysis of a restart algorithm for VC \cite{montanari2002}
is reviewed in Sec.\ \ref{sec:restarts}.

\section{The cov-uncov transition}
\label{sec:transition}

\begin{figure}[ht]
\begin{center}
\scalebox{0.5}{\includegraphics{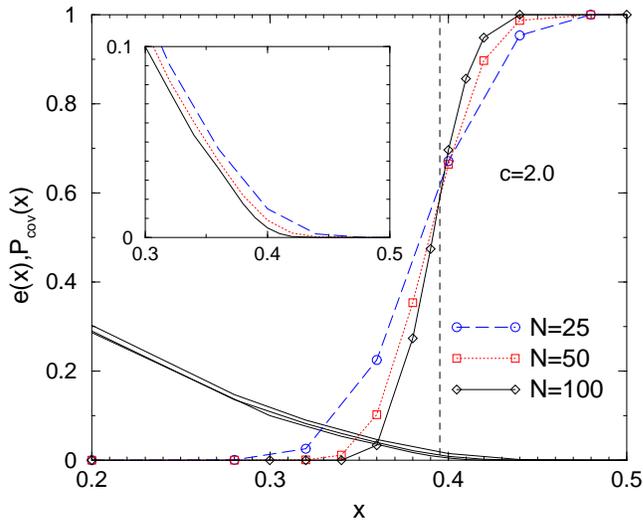}}
\caption{Probability $P_{\rm cov}(x)$  that a cover exists for a
  random graph ($c=2$) as a function of the fraction $x$ of {\em
    covered}\/ vertices. The result is shown for three different
  system sizes $N=25,50,100$ (averaged for $10^3-10^4$ samples). Lines 
  are guides to the eyes only.  In the left part, where the $P_{\rm
    cov}$ is close to zero, the energy average $e$ (see text) is displayed. The
  inset enlarges the result for the energy in the region $0.3\le x \le
  0.5$.}
\label{fig:PEX}
\end{center}
\end{figure}

First, the VC variant is considered where the energy is to be
minimized for fixed values $x=X/N$ (problem P3). We know that for
small values of $x$, the energy density (\ref{eq:coverEnergy}) is not
zero [$e(x=0)=E/N=c/2$], i.e.\ no vertex covers with $xN$ vertices
covered exist.  On the other hand, for large values of $x$, the random
graphs are almost surely coverable, i.e.\ $e(x)=0$. In Fig.\ 
\ref{fig:PEX} the average ground-state energy density and the
probability $P_{\rm cov}(x)$ that a graph is coverable with $xN$
vertices are shown for different system sizes $N=25,50,100$. We consider
her the average connectivity $c=2.0$, but qualitativley
equivalent results are found for other values of $c$ too. The results
\cite{cover,cover-tcs} were obtained using the branch-and-bound
algorithm presented in the last section. The data are averages over
$10^3$ ($N=100$) to $10^4$ ($N=25,50$) samples. As expected, the value
of $P_{\rm cov}(x)$ increases with the fraction of {\em covered}\/
vertices. With growing graph sizes, the curves become steeper. This
indicates that in the limit $N\to\infty$, which we are interested in,
a sharp threshold $x_{\rm c}\approx 0.39$ appears. Above $x_{\rm c}$ a
graph is coverable with probability tending to one in the large-$N$
limit, below $x_c$ it is almost surely uncoverable. Thus, in the
language of a physicist, a {\em phase transition}\/ from an coverable
phase to an uncoverable phase occurs. It is frequently denoted as the
{\it cov-uncov transition}. Note that the value $x_{\rm c}$ of the
critical threshold depends on the average connectivity $c$. The result
for the phase boundary $x_{\rm c}$ as a function of $c$ obtained from
simulations is shown later on.

\begin{figure}[ht]
\begin{center}
\scalebox{0.5}{\includegraphics{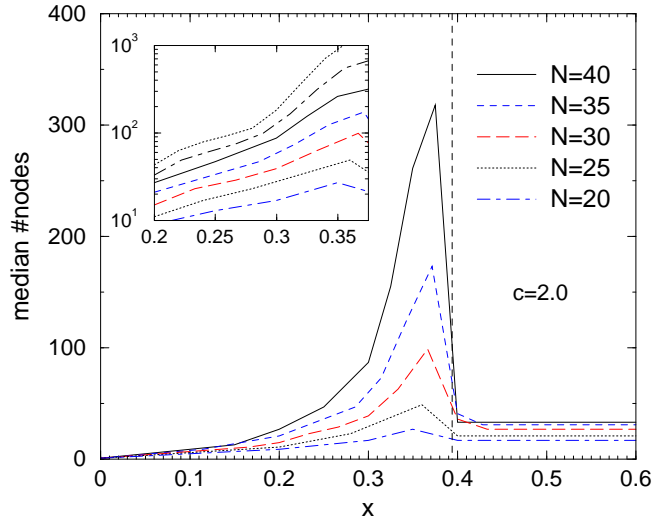}}
\caption{Time complexity of the vertex cover. 
  Median number of nodes visited in the backtracking tree as a
  function of the fraction $x$ of {\em covered}\/ vertices for graph
  sizes $N=20,25,30,35,40$ ($c=2.0$).  The inset shows the region
  below the threshold with logarithmic scale, including also data for
  $N=45,50$. The fact that in this representation the lines are
  equidistant shows that the time complexity grows exponentially with
  $N$.}
\label{fig:TimeX}
\end{center}
\end{figure}

In Fig.\ \ref{fig:TimeX} the median running time of the
branch-and-bound algorithm is shown as a function of the fraction $x$
of covered vertices. The running time is measured in terms of the
number of nodes which are visited in the backtracking tree. Again
graphs with $c=2.0$ were considered and an average over the same
realizations as before has been performed.  A sharp peak can be
observed near the transition $x_{\rm c}$: The hardest instances are
typically found in the vicinity of the phase transition. Note
that also for values $x<x_{\rm c}$ the running time increases
exponentially, as can been seen from the inset of Fig.\ 
\ref{fig:TimeX}. For values $x$ considerably larger than the critical
value $x_{\rm c}$, the running time is linear. The reason is that the
heuristic is already able to find a VC, i.e.\ the algorithm terminates
after the first descent into the backtracking tree\footnote{The
  algorithm used here terminates after a full cover of the graph has
  been found since it is ot necessary to enumerate all solutions}.

\index{vertex cover!time complexity}
\index{time complexity! vertex cover}

Note that continuous phase transitions in physical systems are usually
indicated by a divergence of measurable quantities such as the
specific heat, magnetic susceptibilities or relaxation times. The peak
appearing in the time complexity may be considered as a similar
indicator, but is not really equivalent, because the resolution time
diverges everywhere, only the rate of divergence is much stronger near
the phase transition.

For small values of $x$ in the uncoverable region, the running time is
also faster than near the phase transition, but still exponential.
This is due to the fact that a configuration with a minimum number of
{\em uncovered}\/ edges has to be obtained. If only the question
whether a VC exists or not is to be answered, the algorithm can be
easily improved\footnote{Set $best:=0$ initially.}, such that for
small values of $x$ again a polynomial running time will be obtained.

\section{The phase diagram}

The phase diagram gives the value of the critical threshold $x_{\rm
  c}(c)$ as a function of the connectivity $c$.  For low
connectivities $c<1$ almost all vertices are contained in finite trees
$T_k$ of size $k$ \cite{ErRe,Bo}.  Then one can calculate $x_{\rm
  c}(c)$ using a cluster expansion, i.e. by explicitly calculating
$x_{\rm c}(T_k)$ for small $k$ and weighting the results with the
contribution of each tree $T_k$ to the ensemble of random graphs. In
Ref. \cite{cover-tcs} this expansion has been performed up to tree
size $k=7$, resulting in very good agreement with the numerical data
for small connectivities $c<0.3$.

Using a statistical-mechanics approach it is even possible to derive
an {\em exact} solution, which is furthermore valid even beyond the
percolation threshold $c=1$. We will show that this solution is valid
up to $c=e$, where $e$ is the Eulerian constant. The
statistical-mechanics treatment is presented in the next subsection.
In the second subsection, we will present the results, compare it to
numerical findings and explain the structure of the phase diagram as
well as the solution space structure, finding four different
percolation transitions occurring in VC on random graphs.

\subsection{Mapping VC to a hard-core lattice gas}

To study VC using concepts and methods of statistical mechanics, one
has to map the problem onto a physical system. One possibility is to
identify each vertex with an Ising spin and the two states {\em
  covered/uncovered}\ correspond to the two spin orientations $\pm 1$
\cite{cover}.  Then the system can be studied in the canonical
ensemble and the natural choice for the Hamiltonian is to identify the
energy with the number of uncovered edges (\ref{eq:coverEnergy}).

Here we present a different mapping, using the equivalence between VC
and a hard-core lattice gas \cite{cover-long}.  Any subset $U\subset
V$ of the vertex set can be encoded bijectively as a configuration of
$N$ binary occupation numbers:
\begin{equation}
  \label{eq:mapping}
  x_i := \left\{
      \begin{array}{lll}
       0 & \mbox{if} & i\in U\\
       1 & \mbox{if} & i\notin U
      \end{array}
       \right.
\end{equation}
The strange choice of setting $x_i$ to zero for vertices in $U$
becomes clear if we look to the vertex-cover constraint: An edge is
covered by the elements in $U$ iff at most one of the two end-points
has $x=1$. So the variables $x_i$ can be interpreted as occupation
numbers of vertices by the center of a particle. The covering
constraint translates into a hard sphere constraint for particles of
chemical radius one: If a vertex is occupied, {\it i.e.} $x_i=1$, then
all neighboring vertices have to be empty. We thus introduce a
characteristic function
\begin{equation}
  \label{eq:chi}
  \chi(x_1,...,x_N)=\prod_{\{i,j\}\in E} (1-x_i x_j)
\end{equation}
which equals one whenever $\vec x = (x_1,...,x_N)$ corresponds to a
vertex cover, and zero else. Having in mind this interpretation, we
write down the grand partition function
\begin{equation}
  \label{eq:xi}
  \Xi = \sum_{\{x_i=0,1\}} \exp\left(\mu \sum_i x_i\right) \ \chi(\vec x) 
\end{equation}
with $\mu$ being a chemical potential which can be used to control the
particle number, or the cardinality of $U$. 

For regular lattices, this model is well studied as a lattice model
for the fluid-solid transition, for an overview and the famous
corner-transfer matrix solution of the two-dimensional hard-hexagon
model by Baxter \cite{Ba}. Recently, lattice-gas models with various
kinds of disorder have been considered in connection to glasses
\cite{glass1,glass2,glass3,hclg2003} and granular matter
\cite{granular1,granular2,granular3,granular4,granular5,granular6}.

Denoting the grand canonical average as
\begin{equation}
  \label{eq:mean}
  \langle f(\vec x) \rangle_\mu=\Xi^{-1} \sum_{\{x_i=0,1\}} 
  \exp\left(\mu \sum_i x_i\right) \ \chi(\vec x)\ f(\vec x)
\end{equation}
we can calculate the average occupation density
\begin{equation}
  \label{eq:N}
  \nu(\mu) = \frac{1}{N}\left\langle \sum_i x_i \right\rangle_\mu
  = \frac{\partial}{\partial\mu} \frac{\ln\Xi}{N}\ .
\end{equation}

Minimal vertex covers correspond to densest particle packings.
Considering the weights in (\ref{eq:xi}), it becomes obvious that the
density $\nu(\mu)$ is an increasing function of the chemical potential
$\mu$. Densest packings, or minimal vertex covers, are thus obtained
in the limit $\mu\to\infty$:
\begin{equation}
  \label{eq:mulimit}
  x_c(c) = 1-\lim_{\mu\to\infty} \nu(\mu)\ .
\end{equation}

The main step within the statistical-mechanics approach is to
calculate the grand partition function (\ref{eq:xi}).  Here we state
only the main steps of the calculation without showing intermediate
stage results, details can be found in Ref.\ \cite{cover-long}. The
results of Fig. \ref{fig:PEX} indicate that the model becomes
self-averaging in the thermodynamic limit, i.e. densities of
thermodynamic potentials are expected to become independent on the
specific choice of the quenched disorder (the edge set $E$).
Technically we thus have to calculate the disorder average of the
thermodynamic potential, or the logarithm of the partition function.
The latter can be calculated using the the replica trick
\cite{mezard1987},
\begin{equation}
  \label{eq:rep}
  \overline{\ln \Xi} = \lim_{n\to 0} \frac{\overline{\Xi^n}-1}{n}
\end{equation}
where the over-bar denotes the disorder average over the random-graph
ensemble with fixed average connectivity $c$. Taking $n$ to be a
positive integer at the beginning, the original system is replaced by
$n$ identical copies (including identical edge sets). In this case,
the disorder average is easily obtained, and the $n\to 0$ limit has to
be achieved later by some kind of analytical continuation in $n$.  The
properties of the model can be derived from the $2^n$ order parameters
\cite{monasson1998}
\begin{equation}
  \label{eq:op}
  c(\vec\xi) = \frac{1}{N} \sum_i \prod_a \delta_{\xi^a,x_i^a}
\end{equation}
which give the fraction of vertices having the replicated occupation
number $\vec x_i = \vec\xi\in \{0,1\}^n$. Using this order parameter,
we rewrite the partition function as a functional integral over all
possible normalized distributions $c(\vec\xi)$, $\left(\sum_{\vec\xi}
  c(\vec\xi)=1\right)$. This integral can be evaluated using the
saddle-point method, i.e. one has to optimize over all possible
normalized functions $c(\vec\xi)$. This cannot be performed in full
generality, hence one has to make an ansatz for $c(\vec\xi)$.

The simplest possibility is the so-called replica-symmetric (RS)
ansatz, which in our case assumes that the order parameter
$c(\vec\xi)$ depends on $\vec\xi$ only via $\sum_a\xi_a$, i.e.
different replicas cannot be distinguished, and the full permutation
symmetry of the $n$ replicas is unbroken also on the order-parameter
level. This leads to a specific representation of $c(\vec\xi)$ for
which the replica limit $n\to 0$ can be taken. The resulting
saddle-point equation can now be solved analytically in the limit of
the chemical potential $\mu\to\infty$. The results are presented and
discussed in the next subsection.

Before doing this, let us discuss the validity of the
replica-symmetric ansatz. As it turns out \cite{cover-long} by
considering the local stability of the corresponding saddle-point
solution, this ansatz is valid up to average graph connectivities
$c<e$. At this point full {\em replica symmetry breaking} (RSB)
occurs: Whereas the solution space has a simple geometrical structure
below $c=e$, where all solutions are collected in a single cluster in
configuration space, a hierarchical splitting into many solution
clusters appears continuously at this breaking point.

Despite many efforts, the technical problem of handling RSB in
finite-connectivity systems is still open. Most attempts
\cite{DoMo,MoDo,WoSh,GoLa} try to apply the first step of Parisi's RSB
scheme (1RSB) \cite{mezard1987} which, however, is technically
well-understood only in the case of infinite-connectivity spin
glasses. Due to a more complex structure of the order parameter in
finite connectivity systems, a complete analytical solution is still
missing. Recently, based on the connection to combinatorial
optimization, the interest in this question was renewed
\cite{monasson1998}, and some promising approximation schemes
\cite{monasson1998,brioli2000} have been developed. Even more
recently, a break-through was obtained in context of the cavity method
\cite{MePa2}: Being more involved than the replica method in
infinite-connectivity systems, the cavity approach becomes very
elegant for finite connectivities. It allows for a straight-forward
derivation of self-consistent order-parameter equations at a level,
which is equivalent to 1RSB, and these equations can be efficiently
solved numerically using a population dynamical algorithm. The cavity
method has been recently \cite{zhou2003} applied to VC by Zhou. He
found that, although 1RSB reproduces the numerical results above $c=e$
much better than the replica symmetric solution and satisfies
numerically the bounds presented in Sec. \ref{sec:bound} (see below),
the 1-RSB solution is still not correct above $c=e$. Full RSB has to
be included, which is a completely open technical issue. For this
reason, we refer the reader to Refs. \cite{cover-long,zhou2003} for
the technical details and proceed with the presentation of the
results, mainly for RS.

\subsection{Phase boundary and percolation transitions}

In this section, we describe the analytical results of the statistical
mechanics treatment, compare it to numerical simulations and discuss
the morphology of the phase diagram which can be characterized by the
occurrence of four percolation transitions.

For the density in the limit of infinite chemical potential one
obtains for the RS case
\begin{equation}
  \label{eq:density}
  \nu(\mu\to\infty)
  = \frac{1}{N}\left\langle\sum_i x_i \right\rangle_{\mu\to\infty} 
  = \frac{2W(c)+W(c)^2}{2c}\ ,
\end{equation}
where $W(c)$ is the Lambert-$W$-function \index{Lambert-$W$-function}
defined by
$W(c)\exp(W(c))=c$.
This translates to a minimal vertex-cover size given by
\begin{equation}
  \label{eq:xc}
  x_c(c) = 1- \frac{2W(c)+W(c)^2}{2c}\ .
\end{equation}

\begin{figure}
\begin{center}
\scalebox{0.5}{\includegraphics{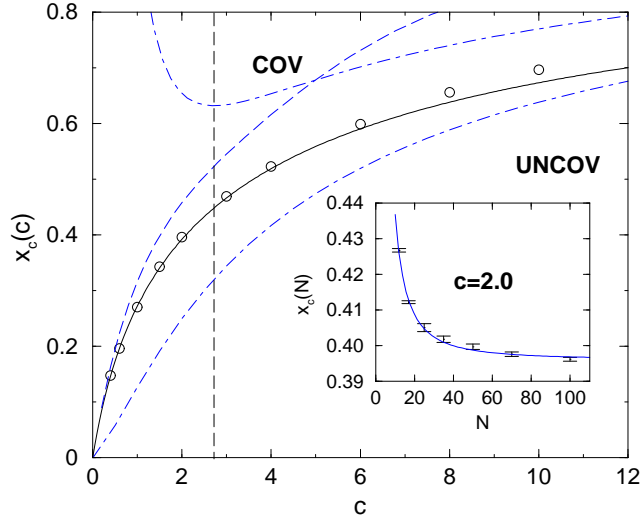}}
\end{center}
\caption{Phase diagram: Fraction $x_{\rm c}(c)$ of vertices in a
  minimal vertex cover as function of the average connectivity $c$.
  For $x>x_{\rm c}(c)$, almost all graphs have vertex covers with $xN$
  vertices, while they have almost surely no cover for $x<x_{\rm
    c}(c)$.  The solid line shows the replica-symmetric result. The
  circles represent the results of numerical simulations. Error bars
  are much smaller than symbol sizes. The upper bound of Harant is
  given by the dashed line, the bounds of Gazmuri by the dash-dotted
  lines.  The vertical line is at $c=e$. Inset: All numerical values
  were calculated from finite-size scaling fits of $x_{\rm c}(N,c)$
  using functions $x_{\rm c}(N)=x_{\rm c}+aN^{-b}$.  We show the data
  for $c=2.0$ as an example.}
\label{figXCCN}
\end{figure}

To calculate the phase boundary numerically, it is sufficient to
construct a single minimal vertex cover. Hence one can apply the
divide-and-conquer algorithm or the version of the branch-and-bound
algorithm where $X$ is not fixed.  To compare with the analytical
results one has to perform the thermodynamic limit $N\to\infty$
numerically. This can be achieved by calculating an average value
$x_{\rm c}(N)$ for different graph sizes $N$, as it is shown for $c=2.0$
in the inset of Fig.\ \ref{figXCCN}. Using the heuristic fit function
\begin{equation}
 x_{\rm c}(N)=x_{\rm c}+aN^{-b} \label{eq:cover:fit} 
\end{equation}
\index{vertex cover!x_c@$x_{\rm c}$} the value of $x_c(\infty)=x_c$
can be estimated from numerical data for finite graphs. As can be seen
from the inset, the fit matches well.

In Fig.\ \ref{figXCCN}, this result is compared to numerical
simulations \cite{cover}. Extremely good coincidence is found for
small connectivities $c<e$.  Up to this value however, we expect the
replica-symmetric result to be exact. This is astonishing, as the
solution does not show any signature of the graph-percolation
transition of the underlying random graph at $c=1$. Please note that
due to the application of statistical mechanics methods like the
replica trick and the replica-symmetric ansatz, the treatment
presented above is not mathematically rigorous. Anyway, for $c<e$, the
result (\ref{eq:xc}) was recently proven to be exact \cite{bauer2001a}
in a constructive way by analyzing a specific VC algorithm. For $c>e$
systematic deviations between the numerical data and the RS estimate
(\ref{eq:xc}) are visible. For large $c$, Eq. (\ref{eq:xc}) even
violates the bounds given in section \ref{sec:bound} and the exactly
known asymptotics (\ref{asympt}), this is due to the appearance of
RSB.

The results \cite{zhou2003} of the cavity-method (corresponding to
1RSB) (not shown) are better than the RS solution since they are
numerically compatible with the asymptotics of \ref{asympt} and within
the bounds of Eqs.\ (\ref{bound_gazmuri}),(\ref{low}). But still the
1RSB solution is significantly different from the numerical
extrapolations in the region $c>e$.

\begin{figure}[htb]
\begin{center}
\myscalebox{\includegraphics{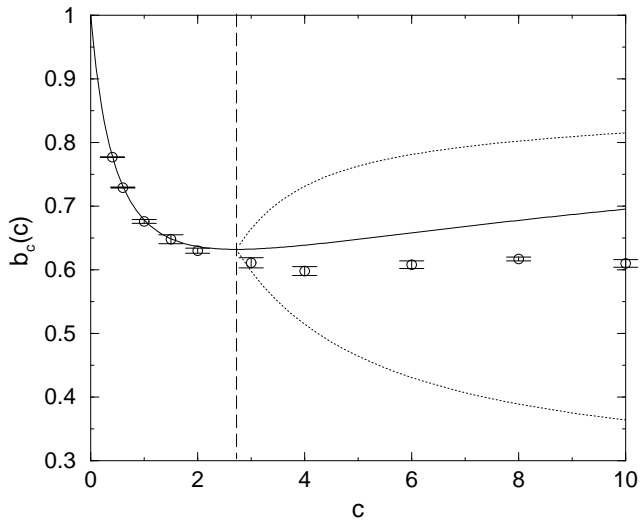}}
\end{center}
\caption{The total backbone size $b_{uncov}(c)+b_{cov}(c)$ of minimal vertex 
covers as a function of $c$. The solid line shows the
replica-symmetric result, the dotted ones are the two results of
one-step RSB. Numerical data are represented by the error 
bars. They were obtained from finite-size scaling fits similar to the 
calculation for $x_c(c)$. The vertical line is at $c=e$ where
replica symmetry breaks down.}
\label{figBCC}
\end{figure}

An important quantity for the understanding of the phase diagram is
the so called {\em backbone}: Usually the minimal VCs are exponentially
numerous. Some vertices are therefore covered in some solutions, but
they are uncovered in other solutions. But there are other vertices
having the same state in {\it all} solutions, being either always
covered or always uncovered. These vertices are frozen in a physical
sense. These vertices are called backbone vertices, we may distinguish
two different types due to the two possible covering states. From the
replica symmetric solution, one can read of the off immediately
\cite{cover-long} the fractions of vertices belonging to these two
backbone types:
\begin{eqnarray}
  \label{eq:bb}
  b_{uncov}(c) &=& \frac{W(c)}{c} \nonumber\\
  b_{cov}(c) &=& 1-\frac{W(c)+W(c)^2}{c} \ .
\end{eqnarray}
The resulting total fraction of backbone vertices of minimum-size VCs 
is shown in Fig. \ref{figBCC}. Numerically, the backbone can be
calculated by enumerating all minimum-size vertex
covers of each realization 
for different sizes $N$ and extrapolating for $N\to\infty$ in a
similar fashion like Eq.\ (\ref{eq:cover:fit}).
For $c<e$ again a very good agreement
is visible. For $c>e$, 
the failure of the RS approach is here even better visible
than when studying the threshold $x_{\rm c}(c)$. Also two results obtained
within the 1-RSB approach (using different ansatzes)
are shown, but they deviate even stronger
from the numerical results.

A detailed analysis \cite{cover-long} shows that vertices having a
small degree are usually uncovered backbone vertices, while the
high-degree vertices usually form the covered backbone. This justifies
a posteriori the use of heuristic algorithm presented in Sec.\ 
\ref{sec:heuristics}.

\begin{figure}[htb]
\begin{center}
\myscaleboxb{\includegraphics{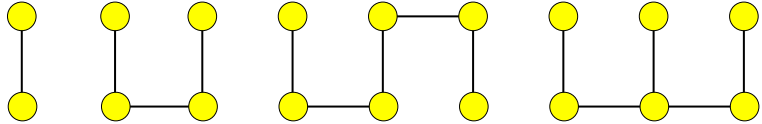}}
\end{center}
\caption{Examples of smallest non-backbone graphs. Note that all this 
graphs can 
be divided into connected vertex pairs and some supplementary edges
connecting different pairs. A similar structure is found also for
the full non-backbone subgraph at connectivities $c<e$.}
\label{fig:nonbb}
\end{figure}

Further results can be obtained when studying the subgraphs induced by
the backbone and the non-backbone \cite{cover-long}. It turns out that
the structure of the non-backbone graphs in the low connectivity
regime $c<e$ can be described as having a collection of pairs, which
are the simplest graphs having no backbone, as building blocks. These
pairs are connected by additional random edges, see e.g.\ Fig.
\ref{fig:nonbb}. The non-backbone subgraphs show a percolation
transition at $c_{\rm bb}=\exp(1/\sqrt{2})/\sqrt{2}$ with
$1<c_{bb}<e$. Hence the onset of RSB at $c=e$ {\em cannot} be
explained by this percolation transition. A similar study for the
backbone subgraphs shows that it percolates already at the original
percolation threshold $c=1$.

Nevertheless, Bauer and Golinelli have indeed related the onset of RSB
to a fourth percolation transition \cite{bauer2001}. They have applied
the leaf-removal algorithm to find minimum-size VCs. The remaining
graph 
is denoted as the {\em
  core} of the graph. Bauer and Golinelli find that, below $c=e$, the
core splits into small disconnected components of logarithmic size,
while above $c=e$ the core percolates and unifies a finite fraction of
all vertices in its largest connected component. Hence, core
percolation seems to be responsible for the onset of RSB!

\section{Analyzing algorithms}

In theoretical computer science the time complexity of an algorithm is
defined as the asymptotic ($N\to\infty$) worst-case running time
measured on a model computer. In real-world applications one is
usually not confronted with this worst case, but with some kind of {\it
  typical} case. As we have seen in Sec.\ \ref{sec:transition} there
might be regions in parameter space (i.e. graph connectivity and VC
size in our case), where VC is typically solved in polynomial time,
while it is typically hard for other parameter regions. Hence, one
would like to observe the easy-hard transition between these regions
within an analytical analysis as well. This would allow for a better
understanding of the underlying mechanisms, hence a step towards
finding the source of computational hardness. We will show that also
here a statistical mechanics treatment, in particular the knowledge of
the phase diagram as calculated before, leads to some interesting
insight.

First, we present the average-case analysis of a simple
branch-and-bound algorithm for the decision problem P2. Within the
algorithm a simple heuristics is used to select the next vertex to
treat.  Next, it is outlined how fluctuations and the influence of
restarts can be incorporated into the analysis. In the third
subsection the analysis of generalized linear-time heuristic
algorithms is summarized.

\subsection{Analysis of a simple branch-and-bound algorithm} 

The algorithm under consideration is a simplified version of the
algorithm presented in Sec.\ \ref{sec:algorithms}. The reason for this
simplification is that it allows for an analytical approach. In the
course of the developments of more sophisticated methods during the
next years which are based on the basic understanding gained for
simple algorithms, it should be possible to analyze more elaborated
algorithms, too.

The simplified branch-and-bound algorithm does {\em not} use the
greedy heuristic, instead the vertices are selected randomly among the
{\em free} vertices. Please note that this corresponds to the case
$w_d=1$ in the generalized heuristic of Sec.\ \ref{sec:heuristics}.
Furthermore the depth $k=0$ is used, i.e. when uncovering a vertex,
its neighbors are not covered immediately. This is also necessary for
simplifying the analysis.  Finally, a simpler bound is used: The
algorithm continues to branch into subtrees as long as covering marks
are available and as long no vertex cover has been found.

The type of analysis presented here was first applied to the 3-SAT
problem by Cocco and Monasson \cite{cocco2001}.  The application to VC
is presented in Ref. \cite{cover-time}.  The analysis of the algorithm
consists of two parts: first, the analysis of the first descent into
the tree and, second, the calculation of full running time, which
includes backtracking if no cover was found in the first descent.  The
running time is measured in terms of the number of nodes visited in
the backtracking tree.

{\em The first descent into the tree:} Previously, probabilistic
analysis of descent algorithms have been applied to establish rigorous
bounds on phase boundaries \cite{ChFr,review1,review2,Ga}. The
analysis of the first descent into the backtracking tree is straight
forward for the algorithm presented here, as it forms a Markov process
of random graphs. In every time step, one vertex and all its incident
edges are covered and can be regarded as removed from the graph. As
the order of appearance of the vertices is not correlated to its
geometrical structure, the graph remains a random graph.  After $T$
steps, we consequently find a graph $G_{N-T,c/N}$ having $N-T$
vertices. As the edge probability remains unchanged, the average
connectivity decreases from c to $(1-T/N)c$.

\begin{figure}[htb]
\begin{center}
\myscalebox{\includegraphics{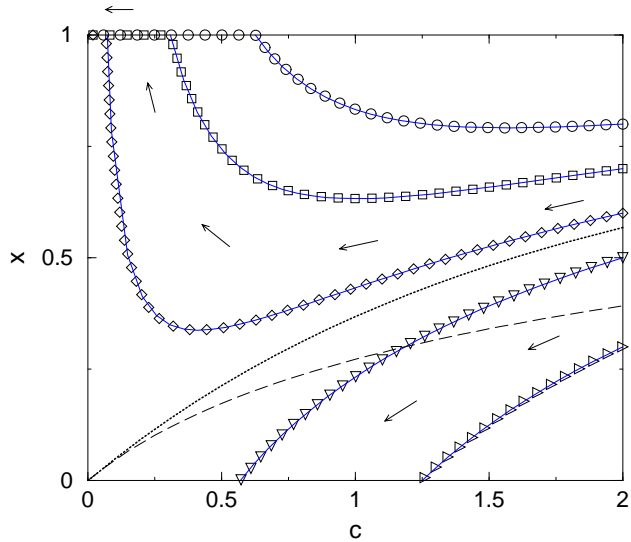}}
\end{center}
\caption{Trajectories of the first descent in the $(c,x)$ plane. The full
  lines represent the analytical curves, the symbols
  numerical results of one random graph with $10^6$ vertices,
  $c=2.0$ and $x=0.8$, $0.7$, $0.6$, $0.5$ and $0.3$. The trajectories
  follow the sense of the arrows. The dotted line $x_b(c)$
  separates the regions where this simple algorithm finds a cover from
  the region where the method fails. No trajectory crosses this
  line.  The long dashed line represents the true phase boundary
  $x_c(c)$, instances below that line are not coverable.}
\label{figTraj}
\end{figure}

For large $N$, it is reasonable to work with the {\it rescaled time}
$t=T/N$, which becomes continuous in the thermodynamic limit. In this
notation, our generated graph reads $G_{(1-t)N,c/N}$. An isolated
vertex is now found with probability $(1-c/N)^{(1-t)N-1}\simeq
\exp\{-(1-t)c\}$, so the expected number of free covering marks
becomes $X(t)=X- N\int_0^t dt^{'}(1- \exp\{-(1-t^{'})c\})$. The first descent
thus describes a trajectory in the $c-x$-plane,
\begin{eqnarray}
  \label{eq:traj}
  c(t) &=& (1-t)c \\ x(t) &=& \frac{x-t}{1-t} +
\frac{e^{-(1-t)c}-e^{-c}}{(1-t)c}. \nonumber
\end{eqnarray}
The results are presented in Fig.\ \ref{figTraj}. One observes a
perfect agreement of the analytical result and the trajectory
generated for a large graph.

\begin{figure}[htb]
\begin{center}
\myscalebox{\includegraphics{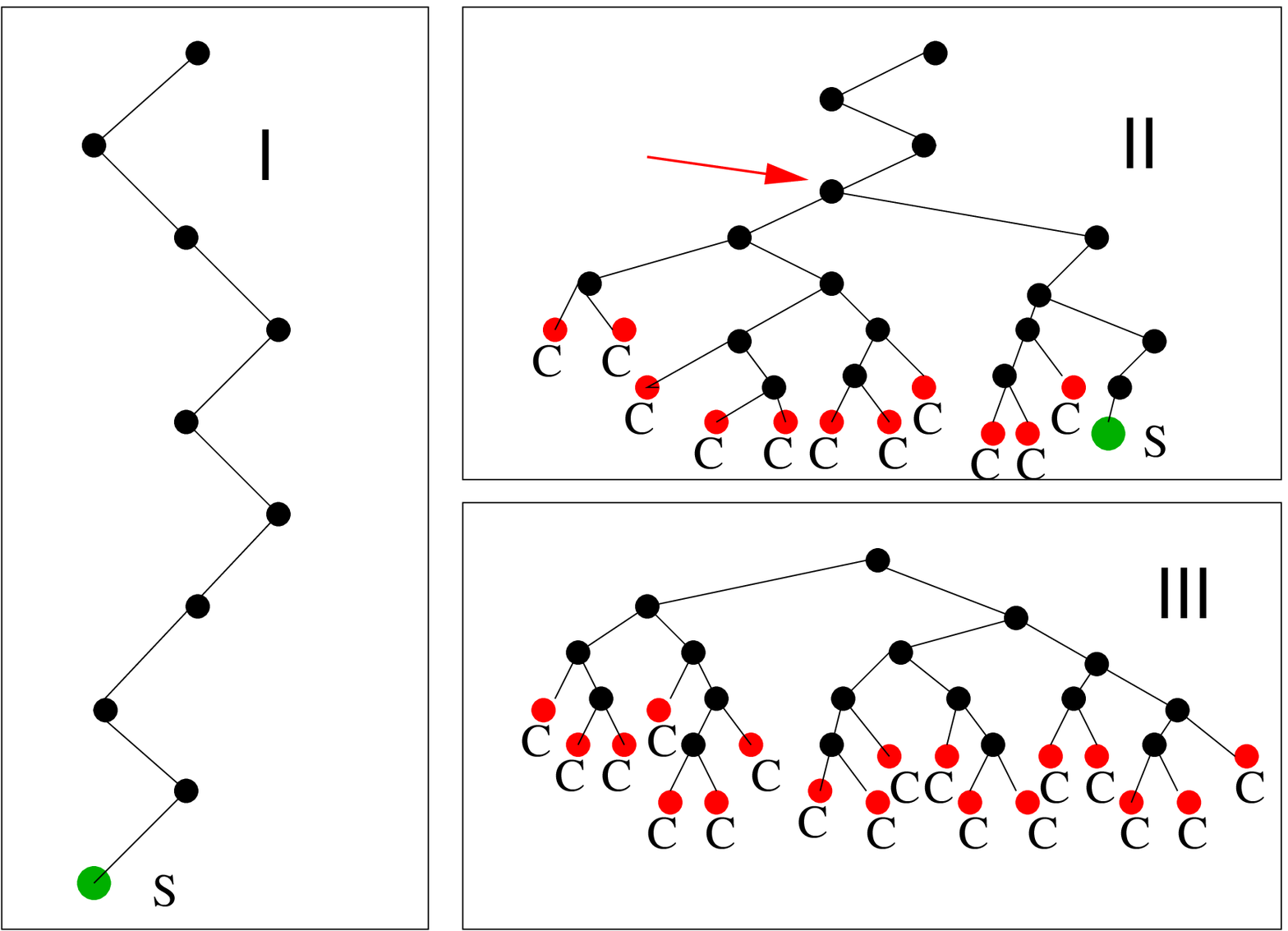}}
\end{center}
\caption{Shape of the backtracking tree in the three dynamical
  regions, contradictions are denoted by ``C'', solutions by ``S'': In
  I, the heuristic immediately finds a solution, no backtracking is
  required. In II, the heuristic fails, the algorithm has to
  backtrack. It has to go back to the tree level, where the first
  uncoverable sub-instance was generated. In III, the graph in
  uncoverable with the given number of covering marks. The algorithm
  has to backtrack completely.}
\label{fig:tree}
\end{figure}

{\em Analysis of the full algorithm:} To understand how the algorithm
works, we study the trajectories together with the phase diagram. We
can observe three regions, the shape of the search tree is
schematically represented in Fig. \ref{fig:tree}:

\begin{itemize}
\item[I] {\it Easy and coverable:} The algorithm works in linear, i.e.
  polynomial time, if the first descent already finds a VC. This is
  the case for large starting value of $x$. Then $x(t)$ reaches the
  value one at a certain rescaled time $t<1$, and the graph is proven
  to be coverable after having visited $tN$ nodes of the backtracking
  tree. The critical value $x_{\rm b}(c)$ above this happens can be
  obtained from (\ref{eq:traj}) by setting $x(t)=1$ and resolving with
  respect to $x$ in the limit $t\to 1$:
  \begin{equation}
    \label{eq:bound}
    x_{\rm b}(c) = 1+\frac{e^{-c}-1}{c}
  \end{equation}
  
\item[II] {\it Hard and coverable:} For $x_{\rm c}(c)<x<x_{\rm b}(c)$
  the graph is typically coverable, but during the first descent
  $x(t)$ vanishes already before having covered all edges. The
  trajectory crosses the phase transition line at a certain rescaled
  time $\tilde{t}$ at $(\tilde{c},\tilde{x})$. There the generated
  random subgraph of $\tilde{N}=(1-\tilde{t})N$ vertices and average
  connectivity $\tilde{c}$ becomes uncoverable by the remaining
  $\tilde{x}\tilde{N}$ covering marks. To determine that the generated
  subproblem is not coverable, the algorithm has basically to visit
  the full backtracking tree for the subproblem. Hence, exponential
  solution times have to be expected. This means $x_{\rm b}(c)>x_{\rm
    c}(c)$ denotes the easy-hard transition of the algorithm.  After
  backtracking the region of the uncoverable subproblem, the
  algorithms proceeds until a solution is found.

\item[III] {\it Hard and uncoverable:} For $x<x_{\rm c}(c)$, the graph
  is typically uncoverable. Thus, again the algorithm has to build a
  full backtracking tree until it is proven that no VC exists. Hence,
  again the running time is exponential.

\end{itemize}

For more sophisticated algorithms, also a phase IV can appear which is
easy and uncoverable. This happens if the used bound is able to prove
already in the very beginning that no VC of the allowed size exists,
and no exponential backtracking is required. The simple algorithm
considered in \cite{cover-time} does not show this phase.

\begin{figure}[htb]
\begin{center}
\myscalebox{\includegraphics{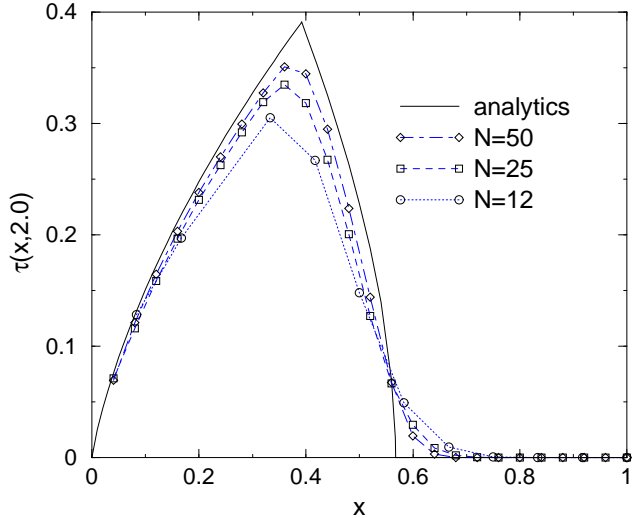}}
\end{center}
\caption{Normalized and averaged logarithm $\tau=\overline{\ln t_{bt}}/N$ of running
  time $t_{bt}$ of the algorithm as a function of the fraction $x$ of coverable
  vertices. The solid line is the result of the annealed calculation. 
  The symbols represent the numerical data for $N=12,25,50$, lines are 
  guide to the eye only. }
\label{figTime}
\end{figure}

To calculate the running time of the algorithm one has to calculate
the size of the backtracking tree generated during the
calculation. This size is determined by the numbers
$\tilde{N},\tilde{c}$ and $\tilde{x}$ characterizing the uncoverable
subproblem which is typically generated. This calculation can be
performed using an annealed approximation and by applying a
saddle-point argument (i.e. the running time is exponentially dominated
by the largest uncoverable subproblem generated). Details can be found
in Ref.\ \cite{cover-time}. The result is displayed in
Fig. \ref{figTime}, where it is compared with numerical simulations.

Note that the algorithm exhibits a peak of the running time exactly at
the phase boundary. This can be directly understood by looking again
at Fig. \ref{figTraj}: For $x>x_{\rm c}(c)$ the uncoverable
subproblems, which have to be backtracked fully, are smaller than the
full graph. For $x<x_{\rm c}(c)$, the number of covering marks is so
small that the generated backtracking trees are smaller due to the
trivial bound included in the algorithm. Thus, directly {\em at} the
phase boundary the size of the backtracking tree is maximal.

\subsection{Fluctuations and random restarts}
\label{sec:restarts}

In the analysis summarized above, the algorithm was assumed to follow
the {\em typical}, or average, trajectory in phase space, and the
generated subproblems become uncoverable exactly when the trajectory
crosses the cov-uncov phase boundary. These assumptions hold with a
probability tending to one in the thermodynamic limit, so they are
perfectly justified if we consider a single run of the algorithm.

There are, however, exponentially rare deviations from these two
assumptions, which can be exploited by running the algorithm described
above only up to some cutoff backtracking time, and restarting it
using a new seed for the random-number generator if no solution was
found. In general we will need exponentially frequent restarts of the
algorithm, but these can be over-compensated by an exponential time
gain due to the restricted backtracking time. According to Montanari
and Zecchina \cite{montanari2002}, the relevant rare events are:
\begin{itemize}
\item Also in the uncoverable phase, there exists an exponentially
  small fraction of coverable instances. Following the first descent
  into the backtracking tree in these rare cases, the system will stay
  coverable up to a point well inside the uncoverable phase. The
  largest generated uncoverable sub-instance will be smaller, and the
  backtracking time consequently shorter. The exponential gain due to
  the smaller backtracking tree has to be balanced against the
  exponential number of restarts needed to find this smaller tree.
  Analytically, these events can be described in a replica calculation
  generalizing the one which was used to calculate the phase boundary.
\item Right from the beginning, the algorithm may follow a different
  trajectory in parameter space, also hitting the phase boundary at a
  later point. Again, macroscopic deviations from the typical
  trajectory are exponentially rare, but they can be exploited by
  exponentially frequent restarts. This can be understood analytically
  within the path-integral formalism introduced by Montanari and
  Zecchina \cite{montanari2002}, by calculating the probability of an
  arbitrary trajectory $(c(t),x(t))$ starting at $(c_0,x_0)$.
\end{itemize}
Most astonishingly, Montanari and Zecchina \cite{montanari2002} found
that the optimal time between restarts is only linear in $N$, i.e.
that mainly no backtracking is needed, because the heuristic is able
to find a solution even in the first descent - even if this happens
with small probability. These analytical results were beautifully
confirmed by numerical simulations. 

In a more general case \cite{gomes2000} this can be different: A
non-trivial optimum in the restart time can be observed numerically
for more sophisticated algorithms.

\subsection{Generalized heuristics}
\label{sec:generalizedHeuristic}

Within the two analysis presented above only a simple heuristic was
considered. The generalized heuristic presented in Sec.\ 
\ref{sec:heuristics} was analyzed by one of the authors
 \cite{weigt2002}, again for
an ensemble of diluted random graphs characterized by an average
connectivity $c$. The concentration of the analysis was laid on the
heuristic itself, not on the interplay with a backtracking algorithm.
The basic idea is similar to the first descent analysis presented in
the preceding section: one follows the dynamics of the algorithm
analytically in a suitably chosen parameter space. For the algorithm
studied in the preceding analysis, the degree distribution $p_d$ of
the graphs is unchanged for all times, i.e it remains the usual random
graph distribution (Poissonian).  Only the average connectivity $c(t)$
is time dependent, leading to a simple differential equation. For the
generalized analysis the degree distribution itself is time dependent,
i.e.  one obtains an infinite set of differential equations for
$p_d(t)$.  The other difference is that in the preceding section the
relative number $x$ of covering marks was given as input to the
algorithm (problem P3), while in this case the algorithms runs until
all edges are covered (problem P1).  The final result of the analysis
gives relative size $x_{\rm f}(c)$ of the resulting VC. This allows to
compare different variants of the heuristic: Algorithms with smaller
$x_{\rm f}(c)$ perform better.

The central idea in the improved heuristic is to select vertices
according to degree-dependent weights $w_d \sim d^\alpha$. This allows,
e.g., for the preferential selection of high-connectivity vertices as
used in the complete algorithm described in Sec. \ref{sec:algorithms}.
In addition, the inclusion of more than one vertex was allowed by
going to depth-$k$ algorithms as already described. The main results
of \cite{weigt2002} are the following:
\begin{itemize}
\item For depth $k=0$, the algorithmic performance increased with
  growing $\alpha$, i.e. with a stronger preference to selecting
  high-connectivity vertices initially. Asymptotically, the
  constructed vertex covers were found to be of size $x_f(c)\simeq 1-
  2\alpha/(c+2\alpha)$. The correct asymptotics of minimal VCs is
  reached to leading order only in the limit $\alpha\to\infty$, which
  is the case implemented in Sec. \ref{sec:algorithms}.
\item For higher depth $k>1$, the correct asymptotics is already
  reached for $\alpha=0$, i.e. for a completely random selection of
  vertices. This includes also the algorithm studied by Gazmuri
  \cite{Ga}, which is characterized by $k=1$ and $\alpha=0$. Still,
  for low connectivities the constructed VCs are pretty large
  compared to the minimal ones.
\item The best performance was found for a generalized leaf-removal
  with $w_d=A\delta_{d,1}+1$. In the limit $A\gg 1$, this algorithm
  unifies the perfect result of leaf removal for small connectivities
  $c<e$ with the correct asymptotic performance of depth-1
  algorithms. 
\end{itemize}
For technical details we refer to \cite{weigt2002}.

\section{VC on other random ensembles}
\label{sec:otherEnsembles}

So far we have presented result for the ensemble of Erd\"os-R\'enyi
random graphs \cite{ErRe}. VC has recently been studied on two other
ensembles, on random graphs with power-law distribution for the
degrees including correlations between vertex degrees, and on graphs
where the basic graph-forming elements are cliques.

V\'azquez and Weigt \cite{vazquez2003} have introduced a generalized
Bethe-Peierls approach, which allows to study VC and other lattice-gas
like models on graphs with arbitrary degree distributions $p_d$.
Furthermore correlations $e_{d,d^\prime}$ between the degrees of
connected vertices are allowed: The quantity $e_{d,d^\prime}$ measures
the probability that for a randomly selected edge, the first
end-vertex has degree $d$, and the second one has degree $d'$.  The RS
result is evaluated for power-law distributions $p_d\sim d^{-\gamma}$
($\gamma >2$) and with correlations
$e_{dd^\prime}=q_d[r\delta_{d,d^\prime}+(1-r)q_{d^\prime} ]$ where
$q_d=(d+1)p_{d+1}/c$ is the probability that for a random edge a
vertex attached to the edge has degree $d+1$. The parameter $r$ can be
used to interpolate between the uncorrelated ($r=0$) regime and the
regime where each vertex is only connected to vertices of the same
degree ($r=1$). The analytical result for the threshold $x_c(r)$ is
compared with numerical results obtained from the application of a
generalized leaf-removal. The leaf-removal process can be used to
determine the onset of RSB: It appears when the number of treated
vertices having minimal degree larger than 1 during the run of the
algorithm becomes of order of the graph size.  The main result is that
for small values of $r$ (e.g. $r<0.7$ for $\gamma=2.5$) the problem is
always easy, i.e. the leaf-removal finds the correct answer. In other
words: Uncorrelated power-law graphs are coverable in polynomial time.
In this region a good coincidence between the analytical and numerical
results could be observed. Results in the RSB region for large $r$ are
not available so far.

A different approach to obtain hard ensembles is followed in Ref.
\cite{hclg2003}. There, graphs are constructed from basic units
consisting of $p$-cliques, i.e.  fully connected subgraphs of $p$
vertices. The full graph is obtained by randomly joining $K$ cliques
in every vertex. VC on such graphs, or the corresponding lattice-gas
model, can be solved using the cavity approach. For $p,K\geq 3$, a
discontinuous 1RSB transition is found at some VC size being
extensively larger than the minimal VC size. This means that the
problem is computationally hard, even if one would be satisfied with a
solution of order $O(1)$ away from the optimum.  Furthermore, when
studying the dynamics using a Monte Carlo algorithm in the
grand-canonical ensemble (see Sec.\ \ref{sec:heuristics}), a dynamical
transition to a glassy phase \cite{Go} appears: The system gets trapped in
metastable states, and equilibration times are exponentially large in
$N$. For this reason, VC on the modified graph ensemble represents a
good mean-field model for structural glass formers. It has, in
particular, only two-particle interactions in contrast to previous
hard-core lattice gas models \cite{FrMeRiWeZe,BiMe,LeDe} for glasses.

\section{Summary and outlook}

We have introduced the vertex-cover problem, which is one of the
fundamental NP-complete problems in theoretical computer science. We
have reviewed different incomplete and complete algorithms for solving
VC. Although VC is considered to be computationally hard, on an
ensemble of random graphs, it exhibits an easy-hard transition when
looking for vertex covers of given size. This make the problem very
valuable for studies aiming for the understanding of the origin of
computational hardness.

Using concepts and methods of statistical physics, many properties of
the model can be understood which are well beyond the horizon of
traditional approaches in theoretical computer science. In the
low-connectivity region ($c<e$, i.e. even above the percolation
threshold $c=1$), it is possible to calculate the phase boundary
exactly using the replica trick or the cavity approach. Above $c=e$
full RSB sets in continuously. The morphology of the phase diagram and
the onset of RSB can be related to different percolation transitions
occurring in the graph and in the solution space structure of vertex
covers.

Furthermore it is possible to analyze analytically simple backtracking
algorithms by following the parameter flow in the phase diagram and to
calculate the easy-hard transition threshold. It is possible to
understand better how an algorithm solves a coverable problem by
including fluctuations in the analysis. Also more complex heuristics,
so far without including backtracking, can be analyzed.

One central point of the future research will be to study special
ensembles of graphs, which are very hard to solve. Examples are graphs
with correlations or graphs having small complete subgraphs. In
particular, one is interested in finite-dimensional regular graphs
(i.e. lattices) exhibiting one-step RSB, which would make them a good
model for structural glass formers.

Another direction of the future research will be the analysis of more
complicated algorithms, i.e. backtracking algorithms with better
heuristics or including bounds. Finally, the research aims to apply
statistical mechanics methods to invent more efficient algorithms,
similar to the recent development of the survey-propagation algorithm
by M\'ezard, Parisi and Zecchina \cite{mezard2002} which originates in
the cavity approach.

\ack

We would like to thank W Barthel, M Leone, R Monasson, F
Ricci-Tersenghi, A V\'azquez, R Zecchina, and H Zhou for fruitful
cooperations and numerous interesting discussions.  Furthermore, we
would like to thank the editors of this special issue, E Marinari, H
Nishimori and F Ricci-Tersenghi, for inviting us to write a review
article. AKH acknowledges financial support from the VolkswagenStifung,
Germany.\\

\end{document}